# Applications of Fission

A.C. Hayes

Theoretical Division, Los Alamos National Laboratory, Los Alamos, NM, 87545, USA

## ABSTRACT

This chapter is devoted to a discussion of applications of nuclear fission. It covers some aspects of the topics of nuclear reactors, nuclear safeguards and non-proliferation, reactor anti-neutrinos and nuclear medicine. It is, however, limited in scope and the reader is encouraged to explore the many other exciting sub-areas of the applications of nuclear fission.

## 1. Introduction

The two key properties of fission, the release of several neutrons and the release of a broad distribution of fission fragments, are fundamental reasons for the huge number of applications that have come since the discovery of the fission process. The importance of neutron emission mainly lies in the uses of fission chain reactions, while the radioactive nature of most of the fission fragments underlies applications from this aspect of fission. The chain reaction nature of fission can be exploited in two ways, controlled fission chain reactions used to generate nuclear energy and uncontrolled chain reactions in explosive nuclear devices. Both of these uses of fission have been in existence since the second world war, when many of the top scientists of the time concentrated on making these applications a reality. Although today the basic and practical principles determining our ability to use fission for nuclear energy or nuclear explosions are well-understood, the initial years required a lot of ingenuity and creativity. Equally creative was the research, mostly started in the 1950s, that went into harnessing the products of fission for medical isotopes. Several fission fragments have ideal decay properties to be used as diagnostics or in radiotherapy to treat some medical conditions, especially cancer. The decay properties of many fission fragments also make them ideal probes for nuclear non-proliferation and reactor safeguards. Another property of the decay of fission fragments is that their beta-decay results in the emission of anti-neutrinos and the anti-neutrino fluxes from nuclear reactors have been used in several worldwide collaborations to examine neutrino oscillations and physics Beyond the Standard Model of particle physics.

The enormous range and diversity of applications of nuclear fission makes it one of the most intriguing sub-atomic processes. In this chapter, I will attempt of discuss a selection of these applications, ranging from nuclear reactors to nuclear non-proliferation to neutrino physics. It is not really possible to cover all of the applications of fission that have been invented by so many resourceful scientific teams. But it is my hope that the discussion of the applications presented here will encourage the reader to explore this rich field in more depth.

## 2. Nuclear Reactors

This section discusses aspects of nuclear reactors that are important for understanding later discussions of non-proliferation and of reactor neutrino physics. In particular, this section pays particular attention to the production of plutonium in reactors. Most modern reactor use low enriched (~2-5% $^{235}$U) uranium fuel and a thermalized neutron flux. The shape of the neutron flux depends on the initial enrichment, and in Fig.1 examples are shown for a pressurized water reactor (PWR) with different enrichments. These fluxes were derived from MCNP/CINDER [1] simulations. With fresh fuel, the burn is initially dominated by fissions from $^{235}$U, but as the burn proceeds, $^{239}$Pu is generated by neutron capture on $^{238}$U, followed by two beta decays, Fig. 2.

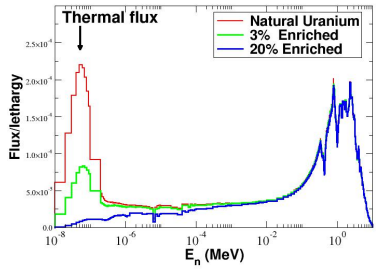

**Fig. 1** *The neutron flux of PWRs with different levels of $^{235}U$ enrichment. The flux is graphed as flux ($\Phi(E)$) per unit lethargy ($\Delta u$), i.e., $\Phi(E)/\Delta u$, where $\Delta u=\ln(E_I/E_{I+1})$.*

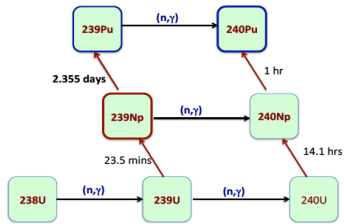

**Fig. 2** *Reactor production of $^{239}Pu$ and $^{240}Pu$ starts with neutron capture on $^{238}U$. The production is complicated by the capture on and beta decay of the intermediate nuclei $^{239}U$, $^{240}U$, $^{239}Np$, and $^{240}Np$. At high neutron flux, $\varphi \sim 10^{14}\,n/cm^2/sec$, the 2.355 half-life of $^{239}Np$ affects the $^{240}Pu/^{239}Pu$ ratio.*

The in-growth rate of the different Pu isotopes depends on the reactor design and the initial fuel enrichment of the uranium fuel. For example, in Fig. 3 we compare the percentage of fissions induced by the different isotopes in the fuel for two PWRs that only differ in the fuel enrichment, one with 2.7% and the other with 4.2% enriched uranium. The total exposure of the 4.2% enriched fuel is about twice that of the 2.7% enriched fuel, and is 60 GW days per metric ton of uranium (GWd/MTU). These simulations were taken from ref. [2] and from earlier coupled calculations with the ERPI/CELL and CINDER-codes [3,4]. Apart from the factor of two difference in total burn times, the two fission histories are very similar. Indeed, increasing the fuel enrichment induces a kind of "accordion" effect in which the *x*-axis is simply stretched.

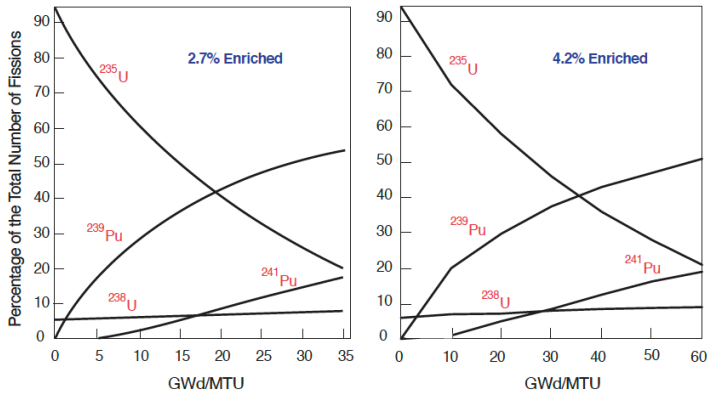

**Fig.3** *The fission history for fresh 2.7% enriched (97.3% $^{238}U$, 2.7% $^{235}U$) and 4.2% enriched (95.8% $^{238}U$, 4.2% $^{235}U$) reactors. As the fuels burn, the importance of $^{235}U$, in terms of the fraction of fissions taking place, decreases steadily. The fission histories are very similar for the two enrichments, apart from the clear difference in burn time.*

## 2.1 Other reactor designs

A number of alternate reactor designs to PWRs exist or have been designed. Perhaps the most famous alternate design is the Canadian design, the CANDU reactor [5]. CANDU designs use natural uranium (0.711% $^{235}U$, 9928% $^{238}U$ and trace $^{234}U$) and are pressurized heavy-water (D₂O) designs. Unlike conventional light-water PWRs, where the core fuel is placed in a pressure vessel, the CANDU bundles of fuel are contained in pressure tubes and these pressure tubes are then contained in a larger unpressurized vessel (a calandria) that is moderated by heavy water. In addition to being capable of producing energy from natural uranium, a large advantage of the CANDU design is that refueling can be done on a continuous basis. This is because only a signal fuel tube needs to be depressurized and replaced at a time. This is in strong contrast to a normal PWR, where the entire core must be shut down in order to open the pressure vessel and refuel the core. There are 30 CANDU reactor operating around the world, in Argentina, Canada, China, India, Pakistan, Romania, and South Korea.

Another very different reactor design is the thorium-uranium fuel design [6]. Thorium is more than 3 time more abundant on Earth than uranium and considerable research has gone into Th-U reactor designs. $^{232}$Th is almost stable with a half-life of 1.405×10$^{10}$ y, and it accounts for 99.98% of the natural thorium on Earth. The remain 0.02% is accounted for by the very small abundance of $^{230}$Th found in parts of the ocean. In general, $^{232}$Th can only be used in breeder reactors because it does not undergo thermal fission. The breeding process is analogous to the production of $^{239}$Pu in uranium reactor, in which $^{232}$Th absorbs a thermal neutron to make $^{233}$Th, and two beta decays result in $^{233}$U, which is fissile.

$$^{232}Th + n \rightarrow {}^{233}Th(22.3 \text{ min}) \xrightarrow{\beta-decay} {}^{233}Pa\ (26.975 \text{ days}) \xrightarrow{\beta-decay} {}^{233}U \quad (1)$$

The thermal neutron capture cross section on $^{232}$Th is larger than that on $^{238}$U, which results in $^{233}$U being bred more efficiently in thorium reactors than plutonium in uranium reactors. To sustain criticality in a thorium reactor normally requires a second fissile material be included in the core to initiate the burn and to provide a neutron flux to allow the nuclear sequence in eq. (1) to proceed. There have been many possible fissile fuels considered for adding to thorium dominated fuel. These have mostly been low-enriched uranium (LEU) or reactor grade plutonium (RGPu). LEU generally refers to uranium enriched up to 20% in $^{235}$U. Higher enrichments would also work, but proliferation becomes an issue. RGPu is generally spent reactor fuel containing about 50% $^{239}$Pu, with the remain Pu coming from $^{238}$Pu, $^{240}$Pu, $^{241}$Pu and $^{242}$Pu.

In Fig. 4 we show the burn history for the case of 80%ThO$_2$+20%LEU. For this example, $^{235}$U initially dominates the fissions, until sufficient $^{233}$U has been bred from the $^{232}$Th. But the $^{238}$U in the LEU fuel also produces 2$^{39}$Pu, which eventually contributes more fission than $^{235}$U.

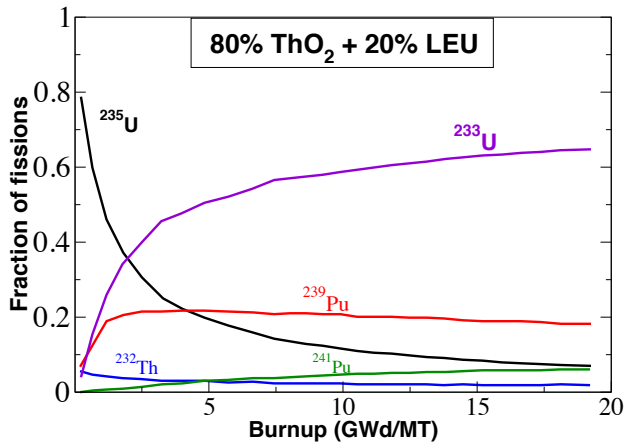

**Fig. 4** *The fission fractions predicted for a Th-U reactor as a function of the burnup. The main role of the $^{232}$Th is to breed $^{233}$U, and $^{233}$U eventually dominates the number of fissions being produced. The LEU fuel is necessary to bring the reactor up to critical and to supply enough neutrons for the breeding of $^{233}$U. The LEU fuel is made up of 2.56% $^{235}$U and 97.44% $^{238}$U. The $^{239}$Pu comes from neutron capture on $^{238}$U, as summarized in Fig.2*

.

**2.2 MOX plutonium fuels**

One of the important concepts for the nuclear fuel cycle is to reuse spent reactor fuel by separating the plutonium and mixing it with depleted uranium to form a Mixed Oxide (MOX) fuel [7]. The plutonium is recovered through reprocessing. Depending on the nation involved in the reprocessing, the spent uranium can also be reprocessed. There are about 40 reactors in Europe (France, Germany Belgium, and Switzerland) that are licensed to use MOX, although only about 30 European reactors are currently using MOX fuel. Japan also recycles MOX fuel in their reactors. Typically, commercial thermal reactors in countries that recycle reactor fuel are loaded with one-third MOX, although some advanced light water designs are capable of accepting 100% MOX loadings.

Another concept that has been explored is to recycle weapons-grade MOX plutonium fuel. The typical isotopic difference between weapons-grade and reactor-grade MOX pluton is summarized in table I

| Isotope % | 235U | 238U | 238Pu | 239Pu | 240Pu | 241Pu | 242Pu |
|---|---|---|---|---|---|---|---|
| Weapons | 0.679 | 93.62 | - | 5.0 | 3.0 | - | - |
| Reactor | 0.679 | 93.62 | 0.203 | 2.653 | 1.373 | 0.535 | 0.535 |

**Table 1.** Initial isotopics for weapons and reactor grade MOX fuels.

There are many MOX loadings that could be considered, but in a detailed analysis we considered [8] four possible cases:
- the entire reactor core is MOX fuel, one case reactor-grade (RG) and another case weapons-grade (WG) plutonium.
- one-third of the core is MOX fuel; the remaining two-thirds of the fuel is assumed to be fresh 2.56% enriched LEU; again, in one case the MOX is RG and in the other case WG.

For all four cases, the Monteburns code [1] was used, which couples the Monte Carlo neutron transport code MCNP [9] to the burn code CINDER'90 [10]; the latter uses 63 neutron energy groups and tracks up to 3400 nuclides, including 638 isomers. In all cases, we assume the same reactor configuration as the H.B. Robinson Unit 2 (HBR2) PWR, which was loaded with a fuel assembly that had 2.56% enriched fresh LEU and no MOX fuel. The simulations reproduce the reported history for the HBR2 PWR using variable concentrations of burnable boron poison rods over the four burn periods and the simulations predict spent fuel isotopic inventories for the uranium and plutonium isotopes within about 5% of measurement. The equivalent burn histories predicted for our four MOX fuel scenarios are shown in Figs. 5 and 6.

As expected, the burn histories for cores involving partial or total MOX fuel are quite distinct from fuel only containing LEU. In the cases where one-third of the core is MOX fuel, the fissions are dominated by $^{235}$U and $^{239}$Pu. The largest difference in the burn between the RG and WG plutonium fuel is the relative importance of $^{241}$Pu. A second difference is the burnup value at which $^{239}$Pu becomes a larger fraction of the fission than $^{235}$U. As is usually the case with all uranium fuel, the relative importance of $^{238}$U remains approximately constant throughout the burn. When the fuel is 100% MOX Pu, $^{239}$Pu dominates the fissions for all burnups. Burning RG fuel differs from WG fuels in the relative importance of $^{239}$Pu versus $^{241}$Pu.

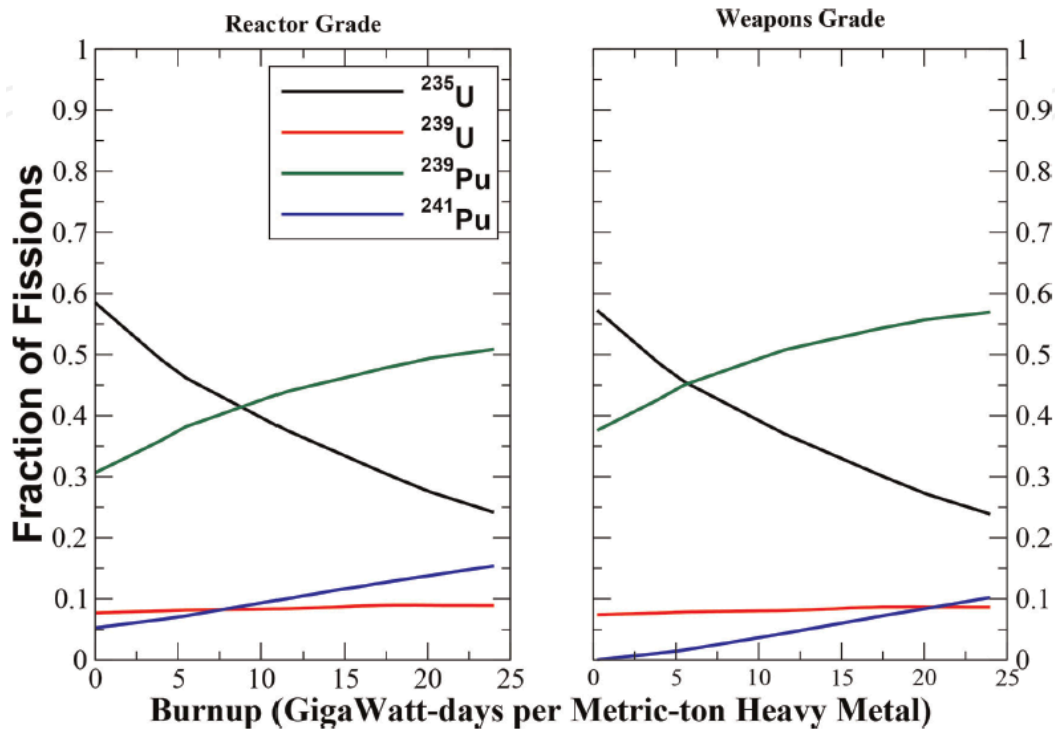

**Fig. 5** Fission *fraction for $^{235,238}$U, and $^{239,241}$Pu as a function of burnup for 33.3% MOX and 66.7%LEU. Panel (a) is RG and (b) WG plutonium.*

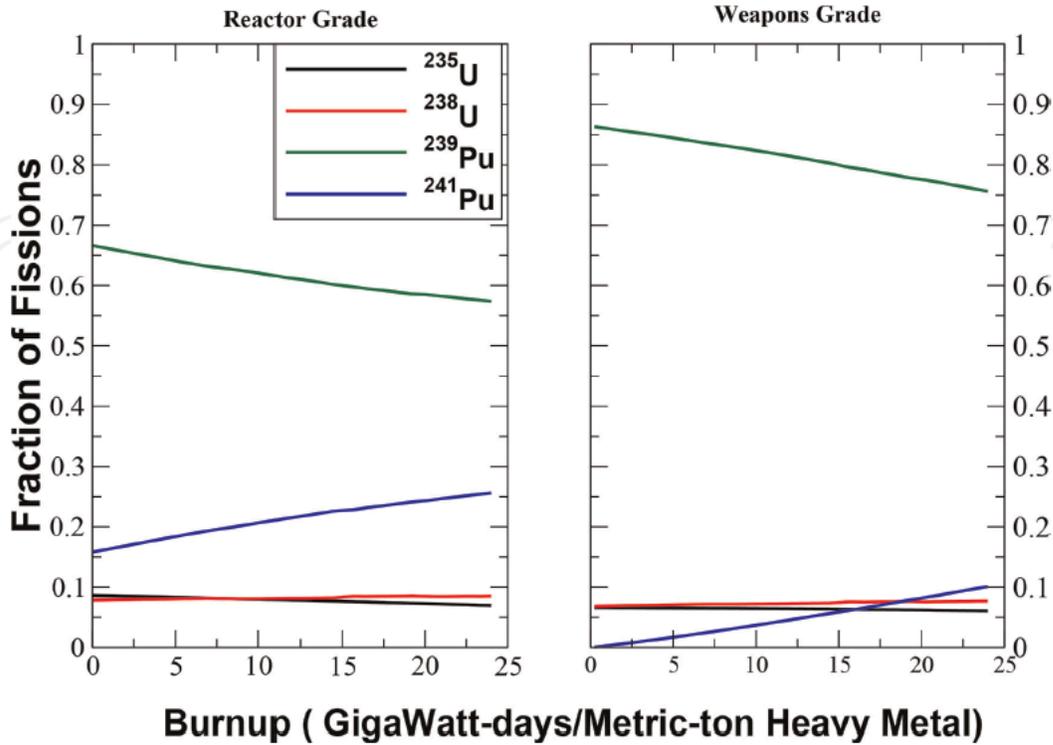

***Fig.6*** *Fission fraction for $^{235,238}$U, and $^{239,241}$Pu as a function of burnup for pure MOX fuels. Panel (a) is for RG and (b) for WG plutonium. The two grades of Pu are distinguished the relative importance of $^{239}$Pu and $^{241}$Pu.*

### 3. Fission Reactor Anti-Neutrinos

One very important application of fission is the use of the anti-neutrinos emitted in the beta-decay of fission fragments. Since the discovery of the anti-neutrino at the Savannah River reactor by Reines and Cowan in 1953 [11-13], neutrino physicists have taken advantage of the intense sources of low-energy anti-neutrinos from nuclear reactors to study the fundamental nature of neutrinos, particularly neutrino oscillations. The concept of neutrinos oscillating [14] from one flavor ($\nu_e$, $\nu_\mu$, $\nu_\tau$) to another is a natural outcome of gauge theories with massive neutrinos and in its simplest form it is expressed as a unitary transformation relating the flavor eigenstates ($\nu_e$, $\nu_\mu$, $\nu_\tau$) to mass ($\nu_1$, $\nu_2$, $\nu_2$) eigenstates.

Understanding the details of anti-neutrino fluxes from reactor is a complex and detailed subject. All of the detectable neutrinos emitted from reactors come from the beta-decay of neutron rich fission fragments, and correspond to an anti-neutrino spectrum ranging from 0-10 MeV. The anti-neutrino spectrum from each fissioning actinides is different and involves thousands of beta-decay branches. The study of neutrino oscillations from reactors has become increasingly important and has now reached a precision science, wherein detailed knowledge of the anti-neutrino spectra is required. This is especially true since the emergence of the so-called reactor neutrino anomaly [15].

The history of reactor neutrino experiments is a very rich one that spans over 60 years. These neutrino oscillation experiments mostly involved searches for variations of the emitted spectrum with the distance from the reactor. In the early experiments, performed in the 1980s-1990s, neutrino detectors were placed

LA-UR-23-22463

within 100 meters of the reactor, but no oscillations were observed. These observations agreed with predictions based on three-neutrino oscillations models and the findings of other neutrino experiments. The first observation of the disappearance of anti-neutrinos from a reactor was made by the KamLAND [16] experiment in the years between 2003-2008. KamLAND detected anti-neutrinos from 56 Japanese reactors at an average distance of about 180 km. By 2012 KamLAND, together with the CHOOZ [17,18] and Palo Verde [19] experiments, had set an upper limit on the neutrino mixing angle $\theta_{13}$ and shown that $\theta_{13}$ was significantly smaller than the mixing angles $\theta_{12}$ and $\theta_{23}$. Following these experiments, the reactor neutrino field moved to much larger detectors and employed both a near detector and a main far detector. The reactor neutrino experiment Daya Bay [20], RENO [21] and Double Chooz [22] determined the value of $\theta_{13}$ with high accuracy. The best fit value from Daya Bay is $\sin^2(2\theta_{13})$=0.084+/-0.005.

### 3.1 The physics determining fission anti-neutrino spectra

For precision studies of neutrino oscillations, accurate knowledge is needed of the fission anti-neutrino spectra for each actinide contributing to the reactor fuel. This issue became a focal point of reactor neutrino studies when in 2011 a reevaluation of the spectra suggested by Huber [23,24] and Mueller [25] (HM) that the expected spectra from all reactors should be increased by about 5-6%. If correct, this would introduce a very serious problem for the neutrino oscillations community because it would mean a need for reassessment of all previous reactor neutrino experiments. In particular, it would imply that the short-baseline experiments were observing neutrino oscillations, but this suggestion could not be accommodated by the standard three neutrino model. New models were proposed to explain the situation, including postulating the existence of a sterile neutrino.

A related problem arose from analyses of the change in the number of anti-neutrinos emitted from reactors with increasing fuel burnup [26]. As summarized in Fig.3, as the burn proceeds in a reactor the relative importance of $^{239}$Pu increases. The total number of anti-neutrinos emitted from the fission of $^{239}$Pu is less than that emitted from $^{235}$U, so as the fuel evolves the number of anti-neutrinos emitted decreases. Daya Bay [26] used the reactor fuel evolution to extract a cross section ratio for $^{235}$U/$^{239}$Pu, $\bar{\sigma}_{235}/\bar{\sigma}_{239}$ =1.445+/-0.097. Here $\bar{\sigma}$ means the average cross section for an anti-neutrino spectrum for the inverse beta-decay process $p + \bar{\nu}_e \rightarrow n + e^+$. However, the HM [23,25] ratio is $\bar{\sigma}_{235}/\bar{\sigma}_{239}$ =1.53+/-0.05. The combination of the reactor neutrino anomaly and the anomalous $\bar{\sigma}_{235}/\bar{\sigma}_{239}$ ratio behooved the nuclear physics community to reexamine the detailed physics determining fission anti-neutrino spectra.

There are two complementary ways to determine the expected anti-neutrino spectrum from a fissioning nucleus [27]: the *ab initio* summation and the electron spectrum conversion methods. In either case, the aggregate fission antineutrino spectrum is determined by summing the contributions of all β-decay branches ($b_{ni}$) of all fission fragments ($Y_n(A_n,Z_n)$),

$$\frac{dN}{dE_{\bar{\nu}}} = \sum_n Y_n(Z_n, A_n) \sum_i b_{ni}(E_0^i) \ S_{\bar{\nu}}(E_{\bar{\nu}}, E_o^i, Z_n) \quad . \quad (2)$$

Here $Y_n(A_n,Z_n)$ is the cumulative fission yield for fragments $(A_n,Z_n)$. The beta-decay spectrum S for a single transition in nucleus (Z, A) with end-point energy E0= Ee + Eν is

$$S(E_e, Z, A) = S_0(E_e) F(E_e, Z, A) C(E_e)(1 + \delta(E_e, A, Z)) \ , \quad (3)$$

where $S_0$= $G_F^2$ $E_e$ $(E_0-E_e)^2$ /$2\pi^3$, $E_e(p_e)$ is the total electron energy (momentum), $F(E_e,Z,A)$ is the Fermi function, and $C(E_e)$ is a shape factor [28] for forbidden transitions. In the case of Gamow-Teller transitions $C(E_e)$=1. The corrections to the spectrum, included in the term δ($E_e$, Z, A), arise from weak magnetism, finite size effects in the Fermi function, and radiative corrections. Changes in the treatment of these corrections was a major source of the initial reactor neutrino anomaly.

In the summation method, one uses the nuclear data bases to carry out the sum over all fission fragments and end-point energies, eq. (2). This method suffers from the problem of the databases being somewhat incomplete. There are theoretical models for some of the missing data, but overall it would be difficult to use the summation method to determine whether an anomaly exists at the 5% level or not.

The second method for determining the anti-neutrino spectra is to convert a measured aggregate beta-spectrum into an anti-neutrino spectrum. This method also comes with difficulties. Any fit to an aggregate beta-spectrum is restricted to about 25-30 (fake) endpoint energies, $E^i_o$. This is because the measured [29-32] aggregate beta spectra are very smooth over 5 orders of magnitude, Fig. 7. Thus, a prescription is needed for the 25-30 values of "$Z_{eff}$" that enter the Fermi functions and this introduces an uncertainty at the few percent level. However, regardless of the prescription used, it can be shown [33] that the ratio $\bar{\sigma}_{235}/\bar{\sigma}_{239}$ is constrained to be close to 1.53+/-0.05, if the Schreckenbach data [29] are used for the conversion.

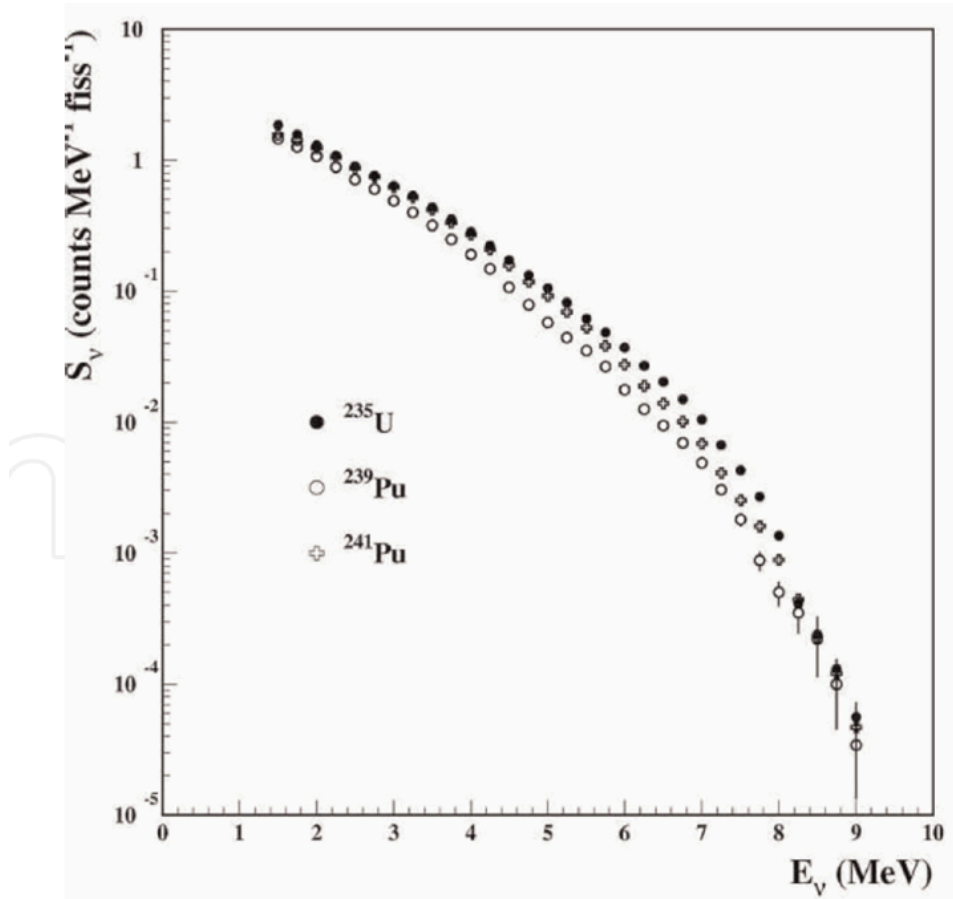

**Fig.7** *The aggregate fission antineutrino spectra for ²³⁵U, ²³⁹Pu and ²⁴¹Pu, deduced from the aggregate beta spectra measured by Schreckenbach et al. [29]. Conversion of these spectra to anti-neutrino spectra by Huber[23] and Mueller[25] led to the so-called reactor neutrino anomaly [15] and to the predicted ratio for ²³⁵U/²³⁹Pu anti-neutrinos inverse beta decay cross sections of 1.53+/-0.05, which is larger than that observed at Daya Bay [26].*

To address this problem, Kopeikin *et al.* [34] remeasured the ratio of the beta spectra for ²³⁵U and ²³⁹Pu at a research reactor in National Research Centre Kurchatov Institute. They found a ratio of $\frac{\bar{\sigma}_{235}}{\bar{\sigma}_{239}} = 1.45 +/-0.03$, which is close to the average value found by Daya Bay and RENO, but 5% lower than any reasonable analysis of the Schreckenbach ²³⁵U and ²³⁹Pu aggregate beta spectra could predict. In addition, the STEREO experiment, which was performed at a reactor with highly enriched ²³⁵U fuel, showed that the anomalous results for the change in the total antineutrino flux with fuel evolutions observed at Daya bay and RENO could be explained entirely in terms of the Schreckenbach measurements having a ²³⁵U spectrum with a normalization that was 5% too high. This then translated into the HM normalization also being too high. It is worth nothing that the predictions of the summation method, using the JEFF-3.1 cumulative fission yields [35] and the ENDF/B-VIII.0 decay data yields [36] yields a ratio $\frac{\bar{\sigma}_{235}}{\bar{\sigma}_{239}} = 1.445$, in agreement with the more modern experiments.

## 4. Nuclear threat reduction and global security

Preventing nuclear proliferation is a world-wide problem that relies on international agreements and on our ability to detect illicit nuclear activities. Of most concern is undeclared production of fissile material, at all stages of any such program. These international efforts include nuclear safeguards for reactors, detecting reprocessing of spent nuclear fuel for the production of weapons plutonium, detecting trafficking of nuclear material, and developing techniques for analyzing debris in the case of a possible terrorist nuclear explosion. The techniques that can be used often depend on the standoff distance. For example, at a compliant reactor, measurements can be made on-site, while for a reprocessing facility in an unfriendly nation, the techniques used may require detection schemes that can work many kilometers from the facility. We consider here some very different situations.

### 4.1 Monitoring the fuel isotopic content in a molten salt reactor

The international community is paying increasing attention to molten salt reactor (MSR) technology as a promising form of clean and reliable energy. In addition to the fuel being dissolved in the salt, many MSR designs involve fuel replacement and reprocessing on a (semi-)continuous basis. These operations render standard techniques, wherein macroscopic samples are removed from fuel rods for assaying the actinide content, useless. Thus, MSR designs raise the challenge of how the isotopic composition of the dissolved fuel content will be determined as the reactor burn progresses. Actinide and fission fragment inventories for standard light water reactors are determined by assaying the spent fuel rods. In the case of MSRs, the fuel is dissolved into the salt so that standard technique cannot be used. In this sub-section we examine a possible method of measuring inventories in-line at the reprocessing station, where the fission gases would be released. The main concept is to replace actinides measurements with actinide *inferences* from the fission gases measurements, applicable to MSRs with continuous refueling and reprocessing operations. Fission gases are automatically released in any reprocessing operation and are easily collected for assay.

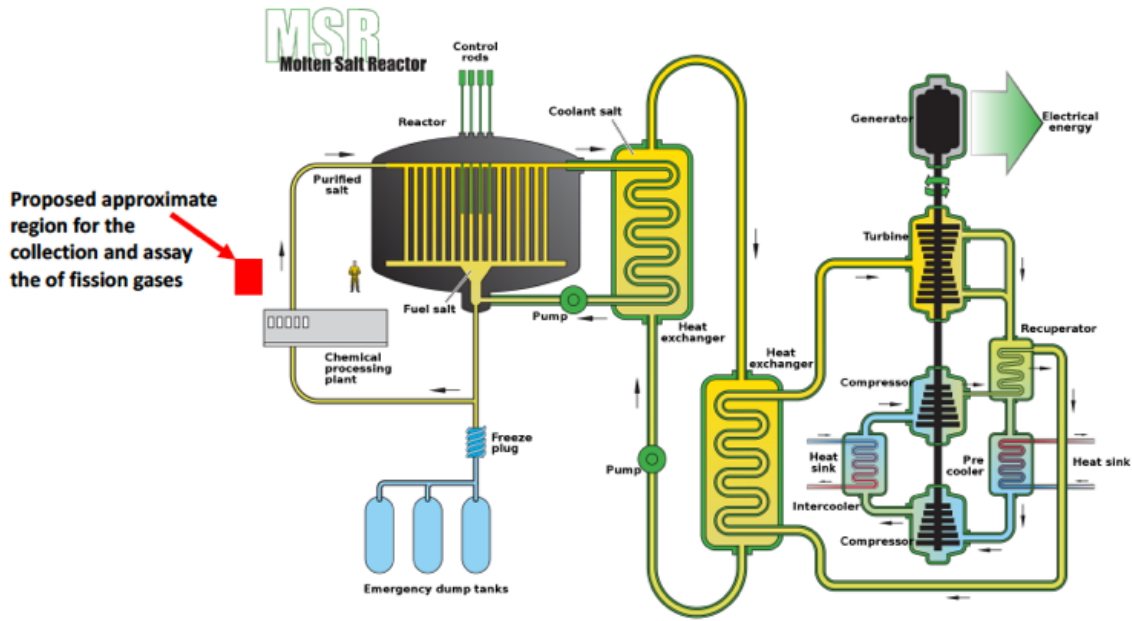

**Fig. 8** *An example of a molten salt reactor, taken from Wikipedia. The fuel is dissolved in the salt. The red box indicates a proposed on-line operation where fission gas measurements would be made, i.e., next to the reprocessing plant.*

In MSR, the salt is typically treated in-line to remove as many fission poisons as possible One set of fission products that are relatively easy to remove are the noble gases, xenon and krypton. This constant separating out of the Xe and Kr gases makes the on-site reprocessing plant an ideal station to monitor the isotopic composition of these gases. As the burn proceeds:
- the stable Kr and Xe isotope abundances (except for $^{136}$Xe, discussed below) simply increase linearly with burnup,
- the unstable Kr and Xe isotopes come to an equilibrium value that is determined by the cumulative fission yield
- the reactor poison $^{135}$Xe asymptotes to an equilibrium value determine by the neutron flux and $^{135}$Xe capture cross section.

So, for example, the growth rate of stable[1] $^{134}$Xe is simply,

$$\frac{dN_{134Xe}}{dt} = \Gamma_f \bar{f}_{134Xe} \quad , \tag{4}$$

where $\Gamma_f$ is the fission rate and $\bar{f}_A$ is the burn-weighted cumulative fission yield of nucleus A. Fig.9 shows the simple linear increase in the $^{134}$Xe/$^{135}$Xe ratio with the burnup for a MSR with Thorium Chloride fuel enriched with 10% $^{233}$U [37]. Thus, a measurement of the $^{134}$Xe/$^{135}$Xe ratio can be inverted to determine the burnup of the fuel, and when coupled with a reactor simulation, the isotopic content.

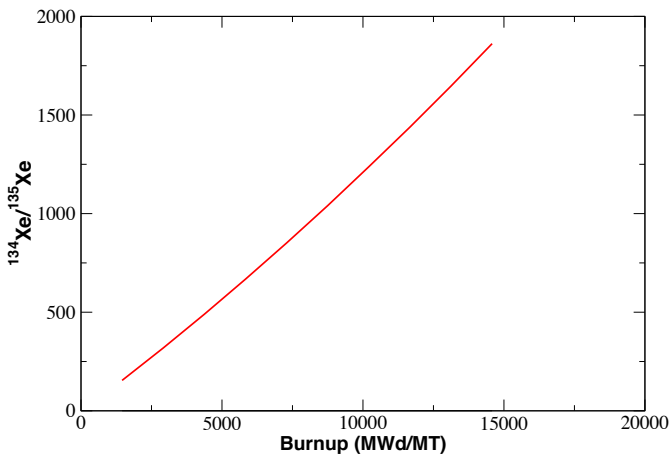

*Fig.9* *The increase in the ratio of $^{134}$Xe/$^{135}$Xe with burnup for a thorium chloride+ 10% $^{233}$U MSR. By measuring the ratio of the stable xenon isotopes to $^{135}$Xe, the fuel burnup, and hence the fuel isotopic content, can be deduced. [from Monteburns simulations [1,37]*

### 4.2 Deducing the grade of plutonium from volatile fission fragments

The grade of plutonium is normally defined to be the ratio $^{240}$Pu/$^{239}$Pu, and weapons grade and reactor-grade fuel are distinguished with the former typically having a $^{240}$Pu/$^{239}$Pu ratio of less than 7% and the latter a ratio of about 25%. If the reactor is running under equilibrium conditions, the $^{240}$Pu/$^{239}$Pu ratio is only dependent on the total neutron fluence, Figs. 10 and 11. There is no dependence on the enrichment of the uranium fuel. However, the relationship between fluence and the $^{240}$Pu/$^{239}$Pu ratio has a small flux dependence, especially at high flux. This is because $^{239}$Pu is produced by the decay of $^{239}$Np, which has a half-life of 2.355 days, and equilibrates at a rate that is flux dependent, Fig. 2.

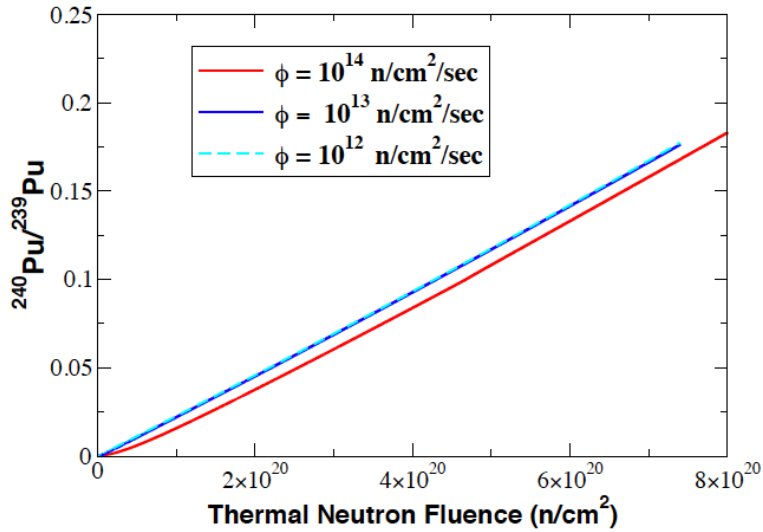

**Fig. 10** *The $^{240/239}$Pu ratio dependence on total neutron fluence. For high flux the curve becomes flux dependent because of the time required for $^{239}$Np to reach equilibrium.*

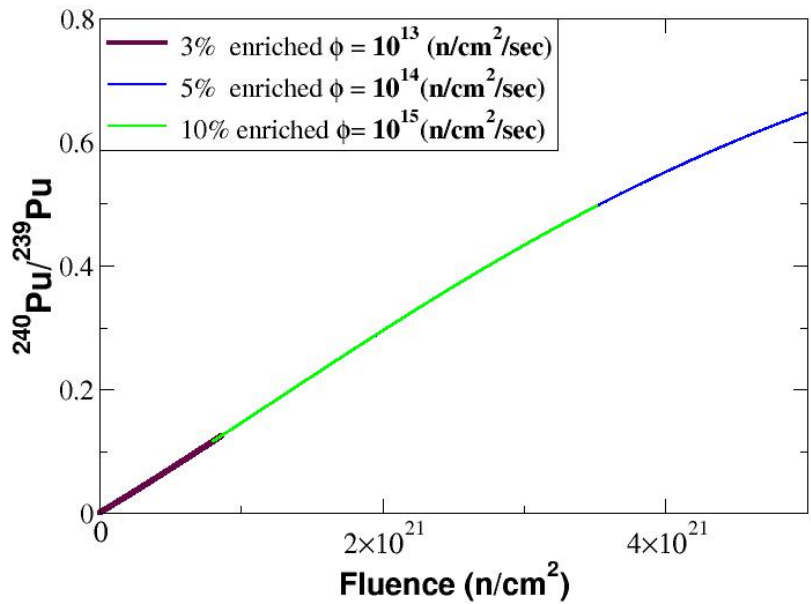

**Fig. 11** *The $^{240/239}$Pu ratio is not dependent on the fuel enrichment. In these simulations, the half-life of 239Np has arbitrarily been set to 1 hour to emphasize the (mostly) simple dependence on fluence.*

Deducing the neutron fluence (n/cm²) that fuel has been exposed to, in order to deduce the grade of plutonium, can be translated into a problem of deducing the neutron flux (n/cm²/sec) and the total irradiation time.

### 4.2.1 Deducing the neutron flux from Xe and Cs fission fragments.

The competition between the 9.14-hour decay of xenon-135 to cesium-135 and the thermal neutron capture on xenon-135 to xenon-136, with an abnormally large cross section of 2.6x10⁶ barns causes the concentration of the fission products $^{136}$Xe and $^{135}$Cs to be sensitive functions of the neutron flux, Fig. 12.

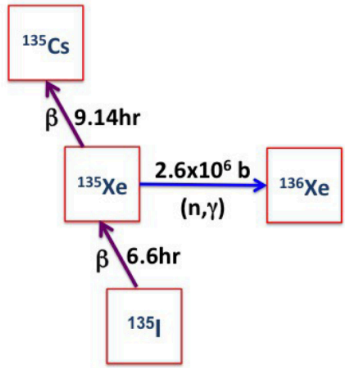

**Fig.12** *The concentrations of $^{136}$Xe and $^{135}$Cs relative to other fission products, such as $^{134}$Xe, are determined by the magnitude of the thermal neutron flux. This sensitivity arises because the rate of thermal neutron capture on $^{135}$Xe competes with rate of decay of $^{135}$Xe to $^{135}$Cs.*

If the thermal neutron flux is low, the $^{135}$Xe dominantly beta-decays to $^{135}$Cs, while if the thermal flux is high, the neutron capture on $^{135}$Xe proceeds faster than the beta-decay because the capture cross section is of the order of 10⁶ barns. In comparison, most fission products are produced directly by fission and/or by the beta-decay of other fission fragments.

$^{135}$Xe is a well-known reactor poison. Under steady state conditions, it reaches an equilibrium value that depends on the capture rate to $^{136}$Xe and on the concentration of $^{135}$I. Changes in the thermal flux cause short-term fluctuations in the level of xenon-135, and after about 40 to 50 hours of steady flux, the xenon-135 level settles to a new equilibrium value that reflects the new flux. Such changes in flux, as well as shutdowns and restarts of the reactor, also affect the $^{136}$Xe/$^{134}$Xe and $^{135}$Cs/$^{137}$Cs ratios.

### 4.2.2 The Dependence of $^{136}$Xe on the reactor thermal neutron flux

In a reactor, $^{136}$Xe is produced by three main mechanisms: as a direct fission product, from the β-decay of $^{136}$I, and from the $^{135}$Xe(n,γ)$^{136}$Xe reaction. The first two mechanisms determine the so-called cumulative fission yield, which is about 7% per fission for both uranium and plutonium.

Under steady state conditions, the growth rate of 136Xe is [38],

$$\dot{N}_{136} = \bar{f}_{136}\Gamma_f + N_{135Xe}^{equil}\phi\sigma$$

$$\dot{N}_{136} = \bar{f}_{136}\Gamma_f \left[1 + \frac{\bar{f}_{135Xe}}{\bar{f}_{136Xe}} \frac{\phi\sigma}{\lambda_{135I} + \phi\sigma}\right] \quad (5)$$

where $\Gamma_f$ is the fission rate, $f_A$ is the burn-weighted cumulative fission yield of nucleus A,
$\phi$ is the thermal neutron flux, $\sigma$ the capture cross section on $^{135}$Xe, and N$^{equil}$ represents the $^{135}$Xe equilibrium value.

The $^{136}$Xe production rate falls between two limits,

$$\phi\sigma \gg \lambda_{135I} : \dot{N}_{136} = \overline{f}_{136}\Gamma_f(1 + \frac{\overline{f}_{135Xe}}{\overline{f}_{136Xe}}) \quad (6)$$

$$\phi\sigma \ll \lambda_{135I} : \dot{N}_{136} = \overline{f}_{136}\Gamma_f$$

The ratio of xenon-136 to xenon-134 is then,

$$\frac{N_{136}}{N_{134}} = \frac{\overline{f}_{136Xe}}{\overline{f}_{134Xe}}\left(1 + \left(\frac{\overline{f}_{135Xe}}{\overline{f}_{134Xe}}\right)\frac{\phi\sigma}{\lambda_{135} + \phi\sigma}\right) \quad (7)$$

From Equation (7), the $^{136}$Xe/$^{134}$Xe ratio equilibrates on a (flux dependent) time scale of weeks. Figure 6 displays the situation for different values of the thermal flux.

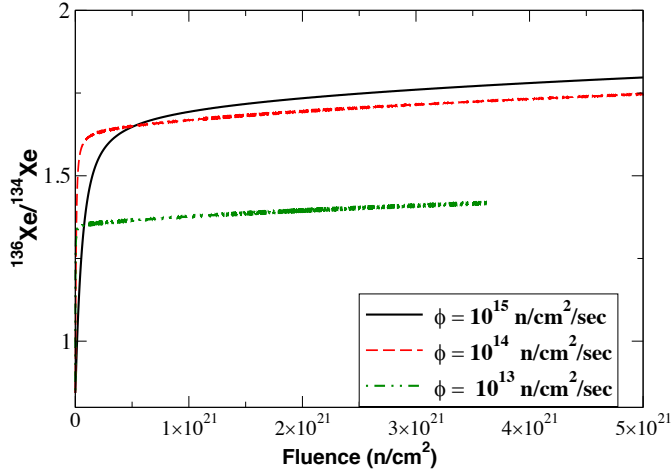

***Fig. 13.*** *The $^{136}Xe/^{134}Xe$ ratio equilibrates to a value that is determined by the thermal neutron flux. Thus, this ratio can be used to deduce the flux. The residual slope of the plateaus in the curves is caused by the slow evolution of the fuel composition, and hence the cumulative fission yields, from [38].*

### 4.2.3 The relationship between thermal flux and the $^{135}$Cs/$^{137}$Cs ratio

In this section we derive an analytic expression relating the concentration of cesium-135 to the thermal neutron flux. Cesium-135 is produced through the decay chain 135I→135Xe→135Cs, which is complicated by the transmutation of xenon-135 via neutron capture. The number of cesium-135 atoms produced in a time t is given by,

$$N_{135Cs}(t) = \lambda_{135Xe}\int_0^t dt' \left[\frac{\frac{\overline{f}_{135I}}{\overline{f}_{137Cs}}\dot{N}_{137Cs} - \dot{N}_{135I} - \dot{N}_{135Xe}}{\lambda_{135Xe} + \phi\sigma}\right] \quad (8)$$

Here $\dot{N}_A$ is the growth rate of nuclide A.

An analytic expression for $^{135}$Cs/$^{137}$Cs ratio can be obtained by assuming that the concentration of iodine-135 and xenon-135 are at their equilibrium values and that the thermal neutron flux is constant. Under these assumptions,

$$N_{135Cs} = \frac{\lambda_{135Xe}}{\lambda_{135Xe} + \phi\sigma}\frac{\overline{f}_{135I}}{\overline{f}_{137Cs}}N_{137Cs} + \frac{\phi\sigma}{\lambda_{135Xe} + \phi\sigma}\left(N_{135I}^{equil} + N_{135IXe}^{equil}\right)$$

$$N_{135Cs} = N_{137Cs}\frac{\overline{f}_{135I}}{\overline{f}_{137Cs}}\left[\frac{\lambda_{135Xe}}{\lambda_{135Xe} + \phi\sigma} + \frac{\phi\sigma}{\lambda_{135Xe} + \phi\sigma}\left(\frac{1}{\lambda_{135I}T_{irrad}}\right)\left(1 + \frac{\lambda_{135I}}{\lambda_{135Xe} + \phi\sigma}\right)\right] \quad (9)$$

Here $\lambda_{135I}$ is the decay constant for iodine-135, and $T_{irrad}$ is the total irradiation time in the reactor. The form of the second term, which gives rise to the term inversely proportional to the total irradiation time, arises from the proper treatment of startup transients.

**4.2.4 The effect of Reactor Shutdowns**

After a reactor shutdown, the operators must wait long enough to ensure that all the $^{135}$Xe and $^{135}$I have decayed before restarting the reactor. Failure to do so was one of the errors that occurred in the Chernobyl accident. Each time the reactor is shutdown, the concentration of $^{135}$Cs is increased. For constant neutron flux operation, shutdown add P terms identical to the second term in equation (9), where P is the number of shutdowns before the fuel is removed. The minimum value of P is one, and the xenon ratio becomes,

$$\frac{N_{136Xe}}{N_{134Xe}} = \frac{\overline{f}_{136Xe}}{\overline{f}_{134Xe}}\left[1 + \frac{\phi\sigma}{\lambda_{135Xe} + \phi\sigma}\left(\frac{\overline{f}_{135Xe}}{\overline{f}_{136Xe}}\right)\left(1 - \frac{P}{\lambda_{135I}T_{irrad}^{total}}\right)\left(1 + \frac{\lambda_{135I}}{\lambda_{135Xe} + \phi\sigma}\right)\right], \quad (10)$$

while the cesium ratio becomes,

$$N_{135Cs} = N_{137Cs}\frac{\overline{f}_{135I}}{\overline{f}_{137Cs}}\left[\frac{\lambda_{135Xe}}{\lambda_{135Xe} + \phi\sigma} + \frac{\phi\sigma}{\lambda_{135Xe} + \phi\sigma}\left(\frac{P}{\lambda_{135I}T_{irrad}^{total}}\right)\left(1 + \frac{\lambda_{135I}}{\lambda_{135Xe} + \phi\sigma}\right)\right]. \quad (11)$$

Here $T_{irrad}^{total}$ the sum of all the irradiation times to which the fuel was exposed. The shutdown correction for the cesium ratio, but not the xenon ratio, becomes significant at high flux. The difference is that for xenon the correction is relative to the constant 1. However, a measurement of the cesium ratio in spent fuel contains information on both the flux and the number of shutdowns.

**4.2.5 Comparisons between the theoretical and experimental cesium and xenon ratios**

Maeck *et al.* measured [39] fission isotope product ratios by irradiating highly enriched uranium targets in the Advanced Test Reactor (ATR) and in the Engineering Test Reactor (ETR) at Idaho National Laboratory. By placing the targets at different distances from the reactor mid-plane, each target was exposed to a different flux, and the flux ranged between 6x10$^{12}$ and 1.5x10$^{14}$ n/cm$^2$/sec. The ATR targets were 93% $^{235}$U and were irradiated for 100 days over 335 days, with several shutdowns. The ETR targets were more than 99% pure $^{235}$U and one set of targets was irradiated for 20 days and another for 180 days over 320 days with some unspecified number of shutdowns.

Figure 14 shows the experimental data and the calculated $^{135}$Cs/$^{137}$Cs ratios. As can be seen, the $^{137}$Cs/$^{135}$Cs ratio scales approximately linearly with the flux, but the reactor shutdowns cause the isotopic ratio to increase with flux at a slower rate at. High fluxes. Figure 15 shows the flux dependence of the $^{136}$Xe/$^{134}$Xe ratio, which is a sensitive diagnostic at low, but not high, fluxes. The $^{136}$Xe/$^{134}$Xe ratio is an ideal probe for low flux graphite reactors, for example.

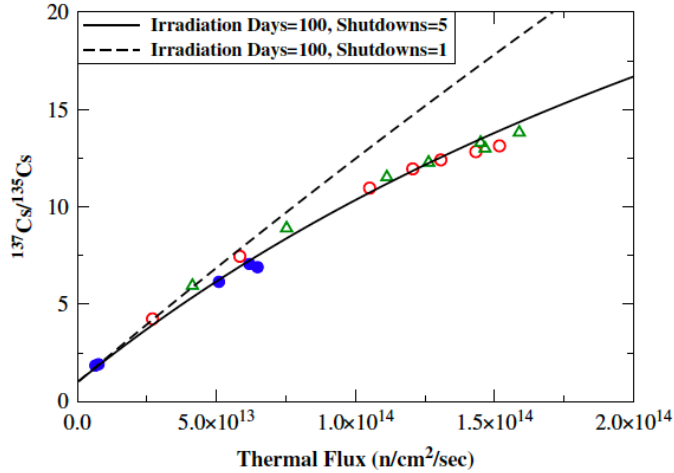

*Fig. 14* The $^{137}$Cs/$^{135}$Cs ratio. The measurements are the data of Maeck et al. [39]. The deviation of the data from the one-shutdown scenario shows the strong dependence of this ratio on the number of shutdowns. The different colors used to display the data points distinguishes the different reactor or irritation times used. The open circles and triangles correspond to targets irradiated at the ATR reactor, while the blue circles are data from targets the ETR reactor.

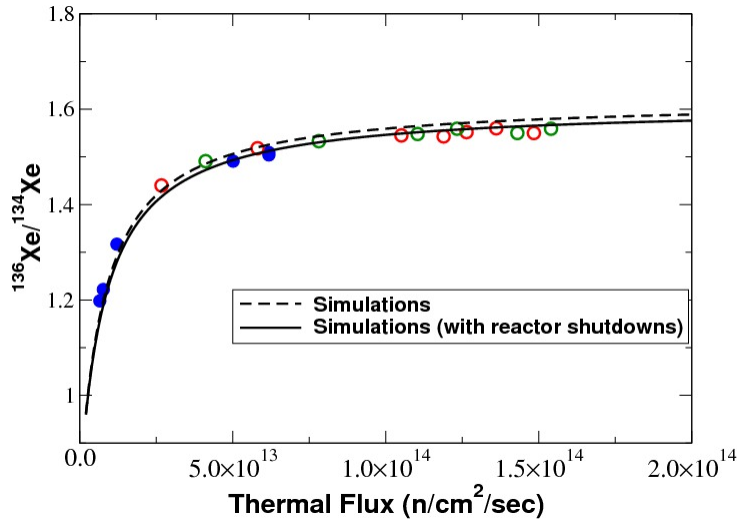

*Fig.15.* The $^{136}$Xe/$^{134}$Xe ratio. The measurements are the data of Maeck [39]. Reactor shutdowns have very little effect on this ratio. The xenon ratio is very good diagnostic for low-flux, but not for high-flux, reactors.

**4.2.5 Monitoring reactor on-off times**

To use the Cs and Xe ratios to address our main problem, namely, the plutonium grade of fuel that is being reprocessed, there remains the problem of knowing the irradiation times and the number of shutdowns. There are several techniques and the usefulness of these depends on the standoff distance from the reactor. We discuss only one method here, namely, reactor neutrino monitoring.

Several neutrino experiments have shown very successfully that measurements of the total number of neutrinos emitted from a reactor can be used to monitor reactor on-off times. Two examples are shown in Fig. 16.

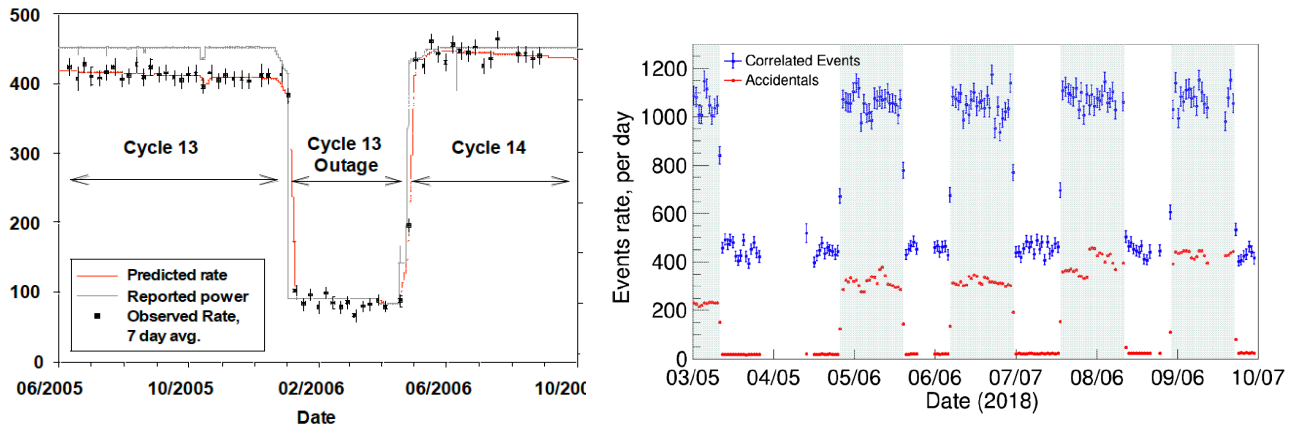

*Fig. 16* Anti-neutrino monitoring of reactors allows of accurate measurements of the reactor on and off burn times. (left) Form the SONGS detector data taken at the San Onofre Reactor [40]. (right) From the PROSPECT detector data taken at the 85 MW HFIR reactor at Oak Ridge [41].

Thus, the grade of reactor fuel can be determined if:
a) The original unirradiated fuel is some form of uranium 238U + 235U, but of unknown enrichment.
b) During the entire burn of the fuel antineutrino monitoring was sufficiently accurate to determine when the reactor was on or off.
c) Fission gases produced in the fuel during the burn can be measured.

This is a nice example of combining two diagnostics to gain enhanced information for non-proliferation, namely, combining reactor fission gas data and antineutrino data.

## 5. Fission applications to nuclear medicine

This section is intended as a short survey of the most common nuclear medical procedures that were derived from fission. In general, uses of radioisotopes and nuclear techniques in medicine fall into two main classes, (a) imaging and diagnosing structures in the body and their functioning, and (b) treating diseases. Fission isotopes contribute to both classes and many reactor facilities around the world are involved in reprocessing spent fuel to produce medical isotopes. The advances in nuclear medicine involve highly multi-disciplinary research and biology and physiological are an important aspect of these studies because of the importance of accurately targeting the organ of interest.

### 5.1 Nuclear Imaging

The $^{99}$Mo-$^{99m}$Tm pair are the most used medical isotope and $^{99m}$Tc is used for imaging and it can be delivered to organs or tissues when injected into the body. Nuclear imaging can detect biochemical changes in an organ (as opposed solely to changes in size), and, thus, can be used to affect decisions in treating disease. The imaging technique used is known as single photon emission computed tomography (SPECT). SPECT [42] uses gamma-rays emitted in the decay of a radionuclide. The $^{9m}$Tc used in SPECT comes from the 2.75-day decay of the fission product $^{99}$Mo. Most of the $^{99}$M is produced in reactors that using highly enriched uranium (HEU) targets, although there is considerable research on moving to production using low-enriched uranium. The isomer of $^{99}$Tc has a half-life of a 6 hours and emits a 140 keV photon. Some fraction of the photons reaches the detector after escaping the body and can be used to create a 3-D image. This is done by taking 2-D projections at different angles. The 3-D dataset that can be manipulated into slices along any axis. One of the advantages of SPECT is that it can be used to study blood flow. The temporal resolution for SPECT is limited, so that only averaged views in time are possible. There has been some discussion in the literature [43,44] about the advantages or disadvantages of SPECT versus Positron Emission Tomography (PET). However, SPECT myocardial perfusion imaging is used almost 20 times more frequently than PET in clinical practices in the United States [44].

## 5.2 Radionuclide Therapy

In targeted radionuclide therapy (TRT) [45], radiopharmaceuticals are delivered to tumor cells by targeting specific characteristics of the tumor, such as antigens. In this way, a toxic level of the radionuclide can be delivered to the site. The radionuclide of choice emits charged particles with the required stopping range for effective destruction of malignant tissue, and include, beta electrons, Auger electrons, and alpha particles. A commonly used fission isotope is $^{90}$Y. As a high-energy beta-emitter $^{90}$Y is used for the treatment of larger tumors and is a successful targeted radiotherapy of the liver. Another application of TRT is in the treatment of bone cancers. A large fraction of breast and prostate cancer patients develop bone metastases, which can cause severe pain. Both $^{153}$Sm and $^{89}$Sr deliver high radiation doses to bone metastases and micro-metastases in the bone marrow. The field of TRT is ever going, with fission fragments playing major role. For example, $^{131}$I, which has traditionally been used to treat thyroid cancer, is now being studied for TRT in metastatic melanoma [46].

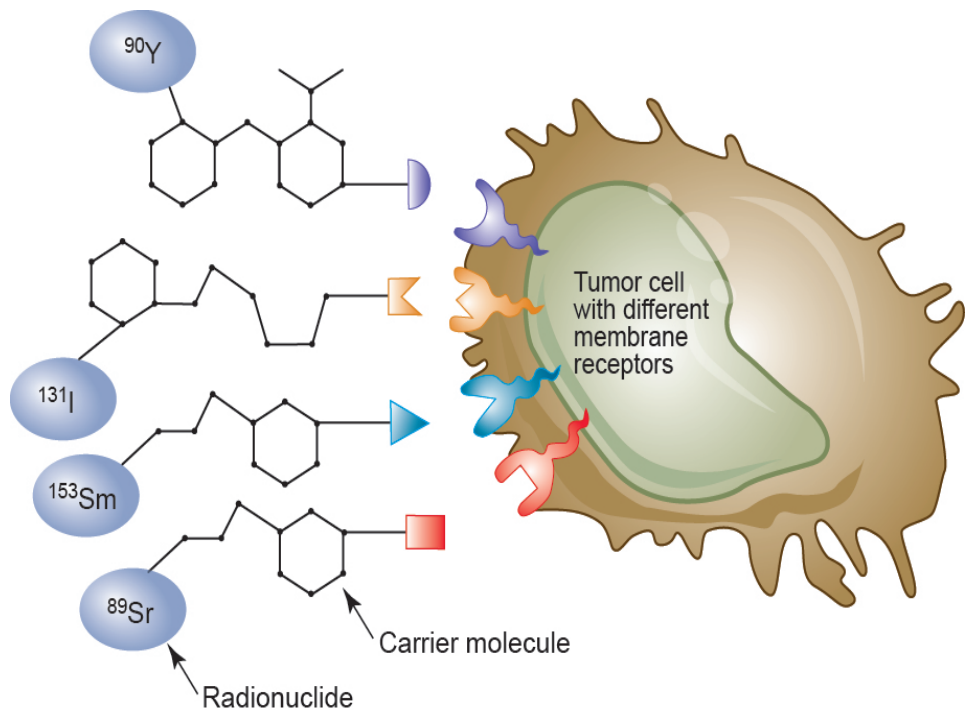

**Fig. 17** *A toxic level of radiation is delivered to a diseased site by attaching a radionuclide to a molecular carrier that binds to the site or tumor.*

## 6 Concluding remarks

Nuclear fission has and continues to be applied to a very broad range of needs. Many of these applications have developed into entire sub-fields of their own and have their own dedicated technical journals. This chapter includes a discussion of nuclear reactors, fission neutrino oscillation studies, nuclear non-proliferation, and nuclear medicine. But there are many more applications of nuclear fission, some of which are discussed in [47]. Nuclear fission is very important for another of other fields, such as nuclear astrophysics, geophysics, nuclear submarines, and nuclear propulsion through space. No doubt the fission of applied nuclear fission will continue to grow as a healthy scientific area.

\